\documentclass[revtex]{emulateapj}
\journalinfo{Accepted for publication in The Astrophysical Journal}

\newcommand{\thetaB}{\theta_{\rm B}}
\newcommand{\thetaopt}{\theta_{\rm opt}}

\shorttitle{Integrated Polarization of Spiral Galaxies}
\shortauthors{Stil et\ al.}

\begin{document}

\title{The integrated polarization of spiral galaxy disks}

\author{Jeroen M. Stil}
\affil{Centre for Radio Astronomy, University of Calgary}
\author{Marita Krause and Rainer Beck} 
\affil{Max Planck Institut f\"ur Radio Astronomie, Auf dem H\"ugel, Bonn, Germany}
\author{A. Russell Taylor}
\affil{Centre for Radio Astronomy, University of Calgary}

\begin{abstract}
We present integrated polarization properties of nearby spiral
galaxies at 4.8 GHz, and models for the integrated polarization of 
spiral galaxy disks as a function of inclination.  Spiral galaxies in
our sample have observed integrated fractional polarization in the
range $\lesssim 1\%$ to $17.6\%$. At inclinations less than 50
degrees, the fractional polarization depends mostly on the ratio of
random to regular magnetic field strength. At higher inclinations,
Faraday depolarization associated with the regular magnetic field
becomes more important.  The observed degree of polarization is lower
($< 4\%$) for more luminous galaxies, in particular those with
$L_{4.8} > 2 \times 10^{21}$ W Hz$^{-1}$. The polarization angle of
the integrated emission is aligned with the apparent minor axis of the
disk for galaxies without a bar. In our axially symmetric models, the
polarization angle of the integrated emission is independent of
wavelength. Simulated distributions of fractional polarization for
randomly oriented spiral galaxies at 4.8 GHz and 1.4 GHz are
presented. We conclude that polarization measurements, e.g. with the
SKA, of unresolved spiral galaxies allow statistical studies of the
magnetic field in disk galaxies using large samples in the local
universe and at high redshift. As these galaxies behave as idealized
background sources without internal Faraday rotation, they can be used
to detect large-scale magnetic fields in the intergalactic medium.
\end{abstract}

\keywords{galaxies: magnetic fields --- galaxies: spiral --- radio continuum: galaxies --- polarization }

\section{Introduction}

Large-scale regular magnetic fields have been detected by Faraday
rotation of radiation of polarized background sources in the Milky Way
\citep{han1999,brown2003,brown2007}, and the nearest galaxies
\citep{stepanov2008,arshakian2007}, such as M31 \citep{han1998}, and
the Large Magellanic Cloud \citep{gaensler2005}. The limiting factor
for this work is the number density of bright polarized background
sources. A few galaxies in the Local Group (M31, LMC) have angular
sizes large enough to probe their magnetic field structure by rotation
measures of polarized background sources with current radio
telescopes. Magnetic fields in galaxies outside the Local Group have
been studied mainly with imaging polarimetry at radio wavelengths
(Beck 2005 for a review). These observations indicate a large-scale
azimuthal magnetic field in the plane of the disk of spiral galaxies,
normally oriented in a spiral pattern with a pitch angle similar to
the optical spiral arms. High degrees of polarization have been
observed in inter-arm regions of some galaxies, such as NGC~6946
\citep{beck2007}. If a magnetic field is observed in a barred galaxy,
it is aligned with the bar inside the region of the bar, with a spiral
pattern outside the bar region \citep{beck2002}. However, some barred
galaxies show no detectable polarized emission or polarized emission
along the bar.

Statistical information on magnetic fields in galaxies is required to
study the cosmic evolution of these magnetic fields by observing
galaxies over a range of redshift. Polarimetric imaging of a large
sample of spiral galaxies {\small in the local universe} is impractical with
current instrumentation because of the high sensitivity and angular
resolution required. On the other hand, deep polarization surveys of
the sky are underway or being planned that will have the sensitivity
to detect spiral galaxies out to large distances, but not resolve
these galaxies.  The first deep wide-area polarimetry survey with the
sensitivity to detect spiral galaxies is
GALFACTS\footnote{http://www.ucalgary.ca/ras/GALFACTS} at 1.4 GHz,
5$\sigma$ point source sensitivity of 400$\mu$Jy, and resolution
$3\arcmin$. SKA pathfinder telescopes such as ASKAP
\citep{johnston2007} aim to do all-sky polarimetry surveys at 1.4 GHz
with an instantaneous band width of 300 MHz with thousands of spectral
channels and a $5\sigma$ point source sensitivity around $50\ \rm \mu
Jy$ \citep{johnston2007}. ASKAP will be able to detect a few thousand
spiral galaxies at sufficient signal to noise ratio to detect
polarization at the few percent level, but the expected angular
resolution of $8\arcsec$ to $\sim 1\arcmin$ will not resolve these
galaxies. This raises the question what can be learned about magnetic
fields from polarimetry of large samples of unresolved spiral
galaxies.

Observations of polarization from spiral galaxies at redshifts up to 1
or higher will require the Square Kilometre
Array\footnote{http://www.skatelescope.org} (SKA), and even then it
will be difficult to make resolved polarization maps of these
high-redshift galaxies. The origin and evolution of cosmic magnetism
is one of five key science drivers for the SKA
\citep{gaensler2004}. For the evolution of magnetic fields in spiral
galaxies as a function of redshift, polarization of the integrated
emission is most readily observable, and a comparison with unresolved
galaxies at low redshift is relatively straightforward.

The number density of faint polarized radio sources in the sky
determines the unknown confusion limit in polarization, and is an
important parameter for rotation measure studies. Below 1 mJy total
flux density, star forming galaxies become an increasingly important
fraction of all radio sources. Knowledge of the polarization
properties of these galaxies is required to make realistic predictions
of the number of polarized sources at $\rm \mu Jy$ levels in polarized
flux density. Spiral galaxies may become an important fraction of the
polarized 1.4 GHz radio source population at $\mu$Jy flux densities
\citep{stil2007a}.

This paper presents observations and models that show that unresolved
spiral galaxies have detectable polarized emission, sometimes more
than 10\% polarized at 4.8 GHz, and that statistical information on
magnetic fields and internal Faraday rotation can be derived from
integrated polarization measurements of spiral galaxies.

\section{Observations}
\label{data-sec}

\begin{deluxetable*}{lrccrcrrrc}
\tablecolumns{10}
\tablewidth{0pc} 
\tabletypesize{\scriptsize}
\tablecaption{ Integrated polarization of nearby spiral galaxies }
\tablehead{
\colhead{Source} & \colhead{Type} & \colhead{$i$} & \colhead{$\nu$} & \colhead{$S_{\nu}$} & \colhead{$S_{\nu}(10\ \rm Mpc)^{a}$} & \colhead{$\Pi_0$} & \colhead{$\thetaB$} & \colhead{$\thetaopt$}  & \colhead{Tel.$^{b}$} \\
\colhead{\ } & \colhead{\ }  &\colhead{($\degr$)}& \colhead{(GHz)} &\colhead{(mJy)} & \colhead{(mJy)} & \colhead{(\%)}  & \colhead{($\degr$)} & \colhead{($\degr$)} &   
}
\startdata
IC 342    & SAB(rs)cd   &   25 &  4.8  & 856   &   82 & 2.2  $\pm$ 0.1 &     50 $\pm$ 1 &      39   &    E        \\
M 31      & SA(s)b      &   76 &  4.8  &1863   &    9 & 17.6 $\pm$ 3.5 &     68 $\pm$ 4 &      38   &    E        \\
M 33      & SA(s)cd     &   56 &  4.8  &1539   &   11 & 1.4  $\pm$ 0.4 &     77 $\pm$ 6 &      23   &    E        \\
M 51      & SAbc        &   20 &  4.8  & 380   &  358 & 2.8  $\pm$ 0.3 &  $-$46 $\pm$ 2 &   $-$10   &    E+V      \\
M 81      & SA(s)ab     &   59 &  4.8  & 385   &   41 & 7.3  $\pm$ 0.3 &  $-$36 $\pm$ 1 &   $-$28   &    E        \\
M 83      & SAB(s)c     &   24 &  4.8  & 809   &  111 & 2.5  $\pm$ 0.4 &      9 $\pm$ 4 &      45   &    E+V      \\
NGC 253   & SAB(s)c     &   78 &  4.8  &2707   &   82 & 1.9  $\pm$ 0.2 &     51 $\pm$ 2 &      52   &    E+V      \\
NGC 891   & SA(s)b? sp  &   88 &  4.8  & 286   &  264 & 1.6  $\pm$ 0.1 &      6 $\pm$ 1 &      23   &    E        \\
NGC 3628  & SAb pec sp  &   89 &  4.8  & 247   &  111 & 4.0  $\pm$ 0.1 &    105 $\pm$ 1 &     104   &    E        \\
NGC 4565  & SA(s)b? sp  &   86 &  4.8  &  54   &   85 & 10.5 $\pm$ 0.3 &  $-$43 $\pm$ 1 &   $-$44   &    E        \\ 
NGC 4736  & (R)SAB(rs)ab &   35 &  4.8  & 125   &   27 & 2.4  $\pm$ 0.4 &  $-$86 $\pm$ 3 &   $-$65   &   E+V      \\ 
NGC 5775  & SAbc        &   81 &  4.8  &  94   &  670 & 0.5  $\pm$ 0.2 &  $-$31 $\pm$ 9 &   $-$35   &    V        \\
NGC 5907  & Sc          &   87 &  4.8  &  72   &   87 & 2.2  $\pm$ 0.3 &   $-$6 $\pm$ 3 &   $-$25   &    E        \\
NGC 6946  & SAB(rs)cd   &   38 &  4.8  & 457   &  224 & 1.5  $\pm$ 0.5 &  $-$82 $\pm$ 7 &      60   &    E+V      \\
\enddata
\tablenotetext{a}{Flux density scaled to a distance of 10 Mpc.}
\tablenotetext{b}{Imaging telescope: E = Effelsberg 100m; E+V = Effelsberg 100m combined with the Very Large Array(VLA).}
\tablenotetext{\ }{References: IC~342: \citet{grave1988}, M~31: \citet{berkhuijsen2003}, M~33: \citet{tabatabaei2007}, 
M~51: \citet{fletcher2004}, M~81: \citet{beck1985}, M~83: \citet{beck2007a} from Beck, Ehle \& Sukumar (unpubl.), 
NGC~253: \citet{heesen2008}, NGC~891, NGC~3628, and NGC~4565: \citet{dumke1998}, NGC~4736: \citet{chyzy_buta2008}, 
NGC~5775: \citet{tullmann2000}, NGC~5907: \citet{dumke2000}, NGC~6946: \citet{beck1996,beck2007}}
\label{nearby-tab}
\end{deluxetable*}

\begin{deluxetable*}{lrccrrrrc}
\tablecolumns{9}
\tablewidth{0pc} 
\tabletypesize{\scriptsize}
\tablecaption{ Integrated polarization of Virgo galaxies }
\tablehead{
\colhead{Source} & \colhead{Type} & \colhead{$i$} & \colhead{$\nu$} & \colhead{$S_{\nu}$} & \colhead{$\Pi_0$} & 
\colhead{$\thetaB$} & \colhead{$\thetaopt$}  & \colhead{Tel.$^{a}$} \\
\colhead{\ } & \colhead{\ }  &\colhead{($\degr$)}& \colhead{(GHz)} &\colhead{(mJy)} & \colhead{(\%)}  & \colhead{($\degr$)} & \colhead{($\degr$)} &   
}
\startdata
NGC 4192  & SABb        &  78    &  8.4  &  16 & 14.5 $\pm$ 1.0 &  $-$28 $\pm$ 1  &     $-$27  &   E      \\
          &             &        &  4.8  &  38 & 11.9 $\pm$ 0.5 &  $-$21 $\pm$ 1  &            &   E      \\
NGC 4254  & SA(s)c      &  42    &  4.8  & 167 &  1.4 $\pm$ 0.1 &  $-46$ $\pm$ 1  &      68    &   E+V    \\
NGC 4302  & SAc         &  90    &  8.4  &  22 &  5.8 $\pm$ 1.0 &  $-$13 $\pm$ 4  &     $-$2   &   E      \\
          &             &        &  4.8  &  36 &  8.4 $\pm$ 0.7 &   $-$8 $\pm$ 1  &            &   E      \\ 
NGC 4303  & SABbc       &  18    &  8.4  & 107 &  1.5 $\pm$ 0.3 &  $-$13 $\pm$ 4  &      $-$28 &   E      \\
          &             &        &  4.8  & 189 &  0.7 $\pm$ 0.1 &  $-$65 $\pm$ 4  &            &   E      \\
NGC 4321  & SAB         &  30    &  8.4  &  66 &  2.3 $\pm$ 0.4 &  $-$58 $\pm$ 4  &      $-$60 &   E      \\
          &             &        &  4.8  & 109 &  1.0 $\pm$ 0.2 &  $-$52 $\pm$ 4  &            &   E      \\
NGC 4388  & SAb         &  82    &  4.8  &  74 &  1.1 $\pm$ 0.1 &  $-$60 $\pm$ 3  &      $-$89 &   E      \\
NGC 4501  & SAb         &  61    &  4.8  & 115 &  1.0 $\pm$ 0.1 &  $-$27 $\pm$ 3  &      $-$42 &   E      \\
NGC 4535  & SABc        &  44    &  8.4  &  20 &  6.7 $\pm$ 1.1 &     17 $\pm$ 3  &      0     &   E      \\
          &             &        &  4.8  &  38 &  8.0 $\pm$ 0.4 &     15 $\pm$ 1  &            &   E      \\
NGC 4654  & SABc        &  56    &  4.8  &  51 &  3.3 $\pm$ 0.2 &  $-$47 $\pm$ 1  &     $-$59  &   E      \\
\enddata
\tablenotetext{a}{Imaging telescope: E = Effelsberg 100m; E+V = Effelsberg 100m combined with the Very Large Array(VLA).}
\tablenotetext{\ }{References: NGC~4192, NGC4302, NGC4303:  unpublished images courtesy M. We\.zgowiec and M. Urbanik,
NGC~4254: \citet{chyzy2007}, NGC~4388, NGC~4501, NGC~4535, and NGC~4654: \citet{wezgowiec2007}.}
\label{virgo-tab}
\end{deluxetable*}

\begin{deluxetable*}{lrccrcrrrc}
\tablecolumns{10}
\tablewidth{0pc} 
\tabletypesize{\scriptsize}
\tablecaption{ Integrated polarization of barred galaxies }
\tablehead{
\colhead{Source} & \colhead{Type} & \colhead{$i$} & \colhead{$\nu$} & \colhead{$S_{\nu}$} & \colhead{$S_{\nu}(10\ \rm Mpc)$} & \colhead{$\Pi_0^{a}$} & \colhead{$\thetaB$} & \colhead{$\thetaopt$}  & \colhead{Tel.$^{b}$} \\
\colhead{\ } & \colhead{\ }  &\colhead{($\degr$)}& \colhead{(GHz)} &\colhead{(mJy)} & \colhead{(mJy)} & \colhead{(\%)}  & \colhead{($\degr$)} & \colhead{($\degr$)} &  
}
\startdata
NGC 1097  & SBbc(rs)    &   45 &  4.8  & 155 & 397 & (0.7 $\pm$ 0.8) &  \ldots &      $-$45  &  V       \\
NGC 1300  & SBb(s)      &   35 &  4.8  &  10 &  38 & 7.6 $\pm$ 2.0\phantom{)} &     89 $\pm$ 6 &      86   &    V       \\
NGC 1313  & SBc(s)      &   38 &  4.8  &  22 &   4 & (0.7 $\pm$ 1.3) &   \ldots &        $-$10  &  A      \\
NGC 1365  & SBb(s)      &   40 &  4.8  & 205 & 740 & (0.3 $\pm$ 0.2) &    \ldots &        40   &    V       \\ 
NGC 1433  & SBb(s)      &   27 &  4.8  &   2 &   3 & (4.7 $\pm$ 3.1) &    \ldots &      17   &    A      \\
NGC 1493  & SBc(rs)     &   30 &  4.8  &   2 &   3 & 13.6 $\pm$ 3.5\phantom{)} &     29 $\pm$ 5 &       \ldots$^{c}$   &    A      \\
NGC 1559  & SBc(s)      &   55 &  4.8  & 105 & 235 &  1.9 $\pm$ 0.1\phantom{)} &  $-$50 $\pm$ 1 &      $-$115 &  A      \\
NGC 1672  & SBb(rs)     &   39 &  4.8  & 105 & 236 &  1.5 $\pm$ 0.2\phantom{)} &  $-$55 $\pm$ 3 &   $-$10  &  A      \\
NGC 2336  & SBbc(r)     &   59 &  4.8  &   4 &  35 &  (3.6 $\pm$ 3.5) &  \ldots &      $-$2   &  V       \\
NGC 2442  & SBbc(rs)    &   24 &  4.8  &  71 & 182 &  1.1 $\pm$ 0.5\phantom{)} &     51 $\pm$ 9 &      40   &    A      \\
NGC 3059  & SBc(s)      &   27 &  4.8  &  29 &  56 &  (0.8 $\pm$ 0.5) &     \ldots &     \ldots$^{c}$   &    A      \\
NGC 3359  & SBc(s)      &   55 &  4.8  &  13 &  29 &  (0.7 $\pm$ 1.4) &     \ldots &      $-$10   & V       \\
NGC 3953  & SBbc(r)     &   61 &  4.8  &   6 &  14 &  8.5 $\pm$ 2.7\phantom{)} &     50 $\pm$ 6 &      13   &    V       \\
NGC 3992  & SBb(rs)     &   59 &  4.8  &   4 &   8 & 14.9 $\pm$ 4.8\phantom{)} &     25 $\pm$ 7 &      67   &    V       \\
NGC 4535  & SBc(s)      &   26 &  4.8  &  12 &  30 &  6.4 $\pm$ 1.4\phantom{)} &      8 $\pm$ 4 &      28   &    V       \\
NGC 4631  & SB(s)d      &   85 &  4.8  & 476 &  268 & 1.1 $\pm$ 0.1\phantom{)} &     16 $\pm$ 2 &      86   &    E        \\  % From nearby sample
NGC 5068  & SBc(s)      &   29 &  4.8  &  14 &   7 &  (1.8 $\pm$ 1.6) &  \ldots &      $-$20&    V       \\
NGC 5643  & SBc(s)      &   30 &  4.8  &  59 & 116 &  1.0 $\pm$ 0.2\phantom{)} &  $-$68 $\pm$ 4 &      \ldots $^{c}$   &    A      \\
NGC 7479  & SBbc(s)     &   45 &  4.8  &  32 & 365 &  3.9 $\pm$ 0.6\phantom{)} &   $-$9 $\pm$ 3 &      25   &    V       \\
NGC 7552  & SBbc(s)     &   31 &  4.8  & 110 & 485 &  1.2 $\pm$ 0.2\phantom{)} &  $-$87 $\pm$ 4 &       1    &    A      \\
\enddata
\tablenotetext{a}{Galaxies with $\Pi_0$ less than 2 times the formal error listed in parentheses; these galaxies are not included in figures}
\tablenotetext{b}{Imaging telescope: A = Autralia Telescope Compact Array (ATCA), E =  Effelsberg 100 m, V = VLA.}
\tablenotetext{c}{Position angle of major axis undetermined for this low-inclination galaxy.}
\tablenotetext{\ } {References: NGC~4631: Krause et al. (unpubl.) but see \citet{krause2003} and \citet{golla1994}; all other galaxies from \citet{beck2002}.}
\label{barred-tab}
\end{deluxetable*}

We use a compilation of archival data for nearby galaxies imaged in
polarized radio emission.  Observations at 4.8 GHz are most widely
available, so this paper focuses at this frequency to maximize the
sample size for a single frequency. Additional observations at 8.4 GHz
are available for five galaxies. These are listed alongside the 4.8
GHz data.  Three sub samples are distinguished in this paper. The first
sub sample is a set of 14 nearby ``field'' galaxies listed in
Table~\ref{nearby-tab}. The second sub sample is a set of 9 spiral
galaxies in the Virgo cluster from \citet{wezgowiec2007}, We\.zgowiec
et~al. (in prep.), and \citet{chyzy2007} listed in
Table~\ref{virgo-tab}. These galaxies exist in a cluster environment,
with a higher probability of interaction with other cluster members
and the intracluster gas. The third sub sample of 20 barred spiral
galaxies listed in Table~\ref{barred-tab} allows us to explore the
integrated polarization of barred galaxies. One nearby galaxy
(NGC~4631, classified as SB) was included in Table~\ref{barred-tab}.

The nearest galaxies in the sample and the galaxies in the Virgo
cluster were observed with the 100m-Effelsberg telescope, sometimes in
combination with the VLA. These galaxies are not affected by missing
short spacings in the interferometer. The barred galaxies were not
observed with the Effelsberg telescope, and the images could in
principle lack large scale structure resolved by the shortest
baselines of the interferometer. A comparison of 6 cm flux densities
derived by \citet{beck2002} with published integrated flux densities
showed that NGC~1313 and NGC~3992 in Table~\ref{barred-tab} may be
affected by missing short spacings, because published flux densities
listed in the NASA Extragalactic Database (NED) are substantially
higher than the 6-cm flux density in \citet{beck2002}. If no 6 cm flux
density was published, we compared the 21 cm flux density corrected
for a spectral index 0.7 - 1 with the flux density in
\citet{beck2002}. Of the two barred galaxies with suspected missing
flux in the interferometer image, NGC~1313 showed no significant
integrated polarization, while NGC~3992 has a high degree of
polarization of $14.9 \pm 4.8\ \%$. Most of the emission of NGC~3992
detected in the interferometer image originates outside the bar region
\citep{beck2002}.

\begin{figure*}
\resizebox{\textwidth}{!}{\includegraphics[angle=0]{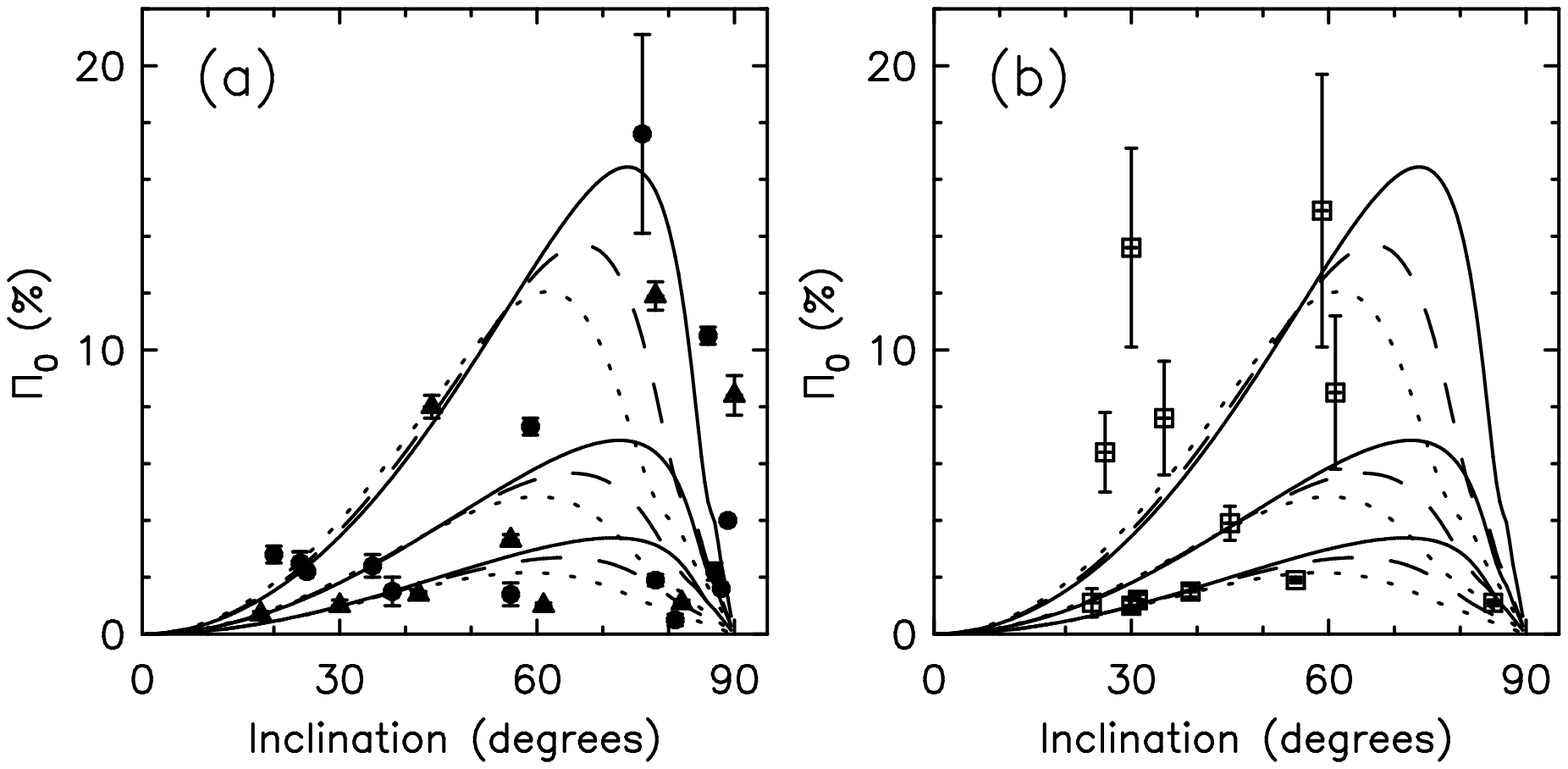}}
\caption{ (a) Integrated fractional polarization at 4.8 GHz of nearby
spiral galaxies from Table~\ref{nearby-tab} (filled circles) and Virgo
cluster spirals from Table~\ref{virgo-tab} (triangles), with model
curves described in Section~\ref{model-sec}. The model curves from top
to bottom correspond with models 1--9 in Table~\ref{model_par-tab}, in
that order. (b) The same as (a) for the detected barred galaxies from
Table~\ref{barred-tab} (squares). The model curves in (b) are shown
for reference only.
\label{diskpol_model-fig}
}  
\end{figure*}

\begin{figure*}
\resizebox{8.5cm}{!}{\includegraphics[angle=0]{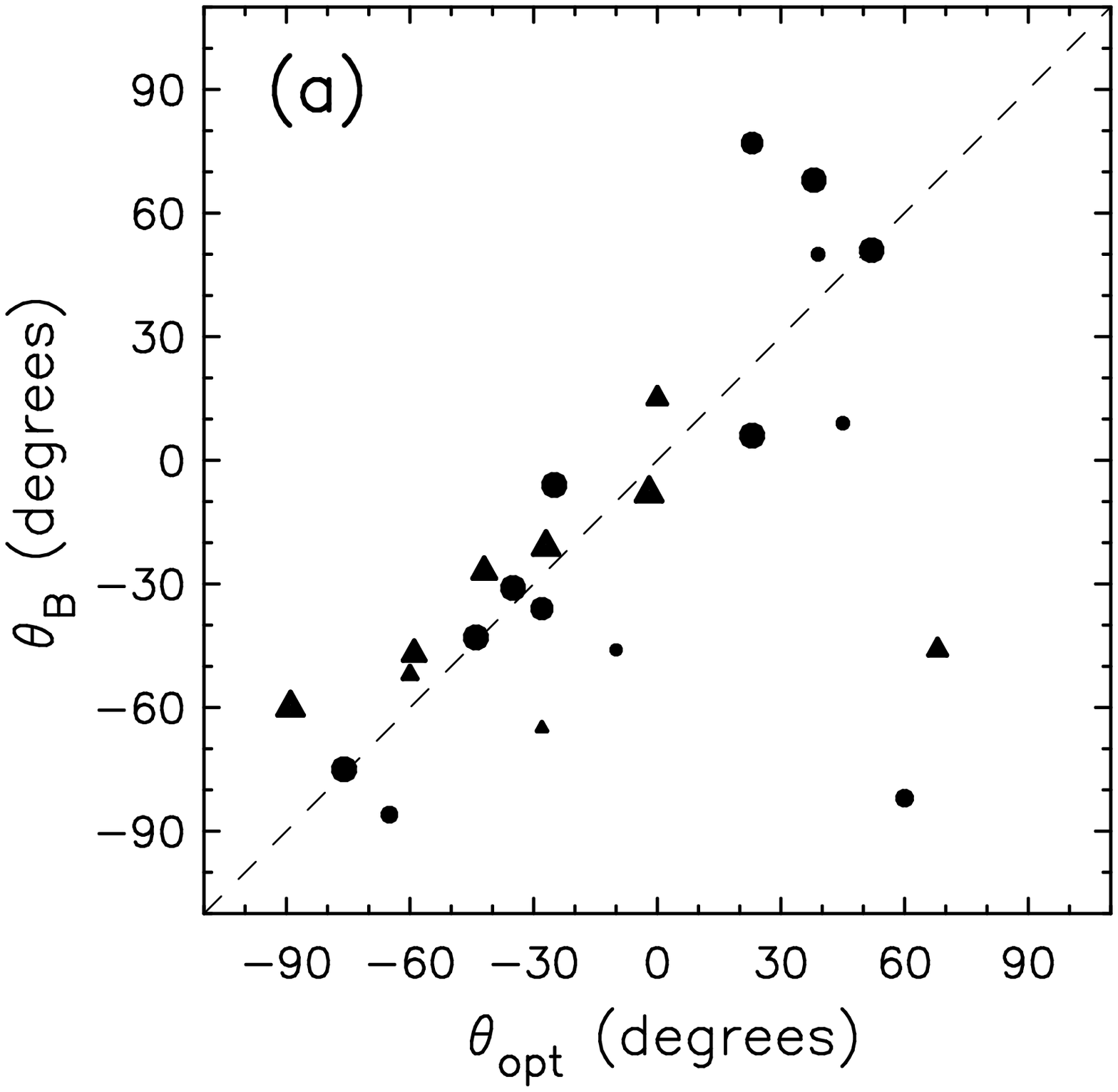}}
\resizebox{8.5cm}{!}{\includegraphics[angle=0]{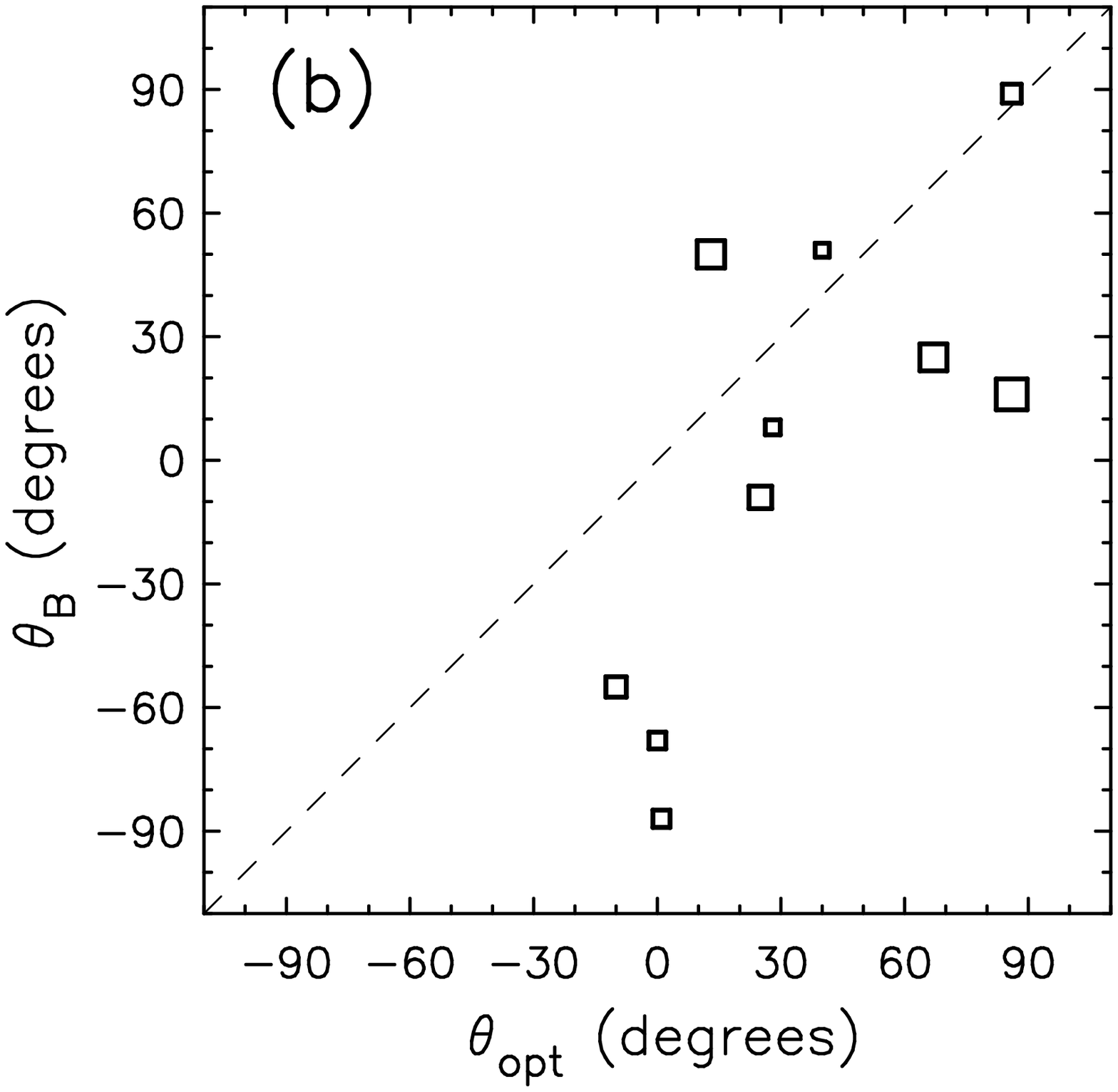}}
\caption{ (a) Correlation between $\thetaopt$ and $\thetaB$ for
galaxies in Table~\ref{nearby-tab} (filled circles) and the Virgo
galaxies in Table~\ref{virgo-tab} (triangles).  The symbol size
increases linearly with $\sin(i)$. (b) The same as (a) for the barred
galaxies in Table~\ref{barred-tab} (squares).
\label{PA-fig}
}  
\end{figure*}

\begin{figure*}
\resizebox{8.5cm}{!}{\includegraphics[angle=0]{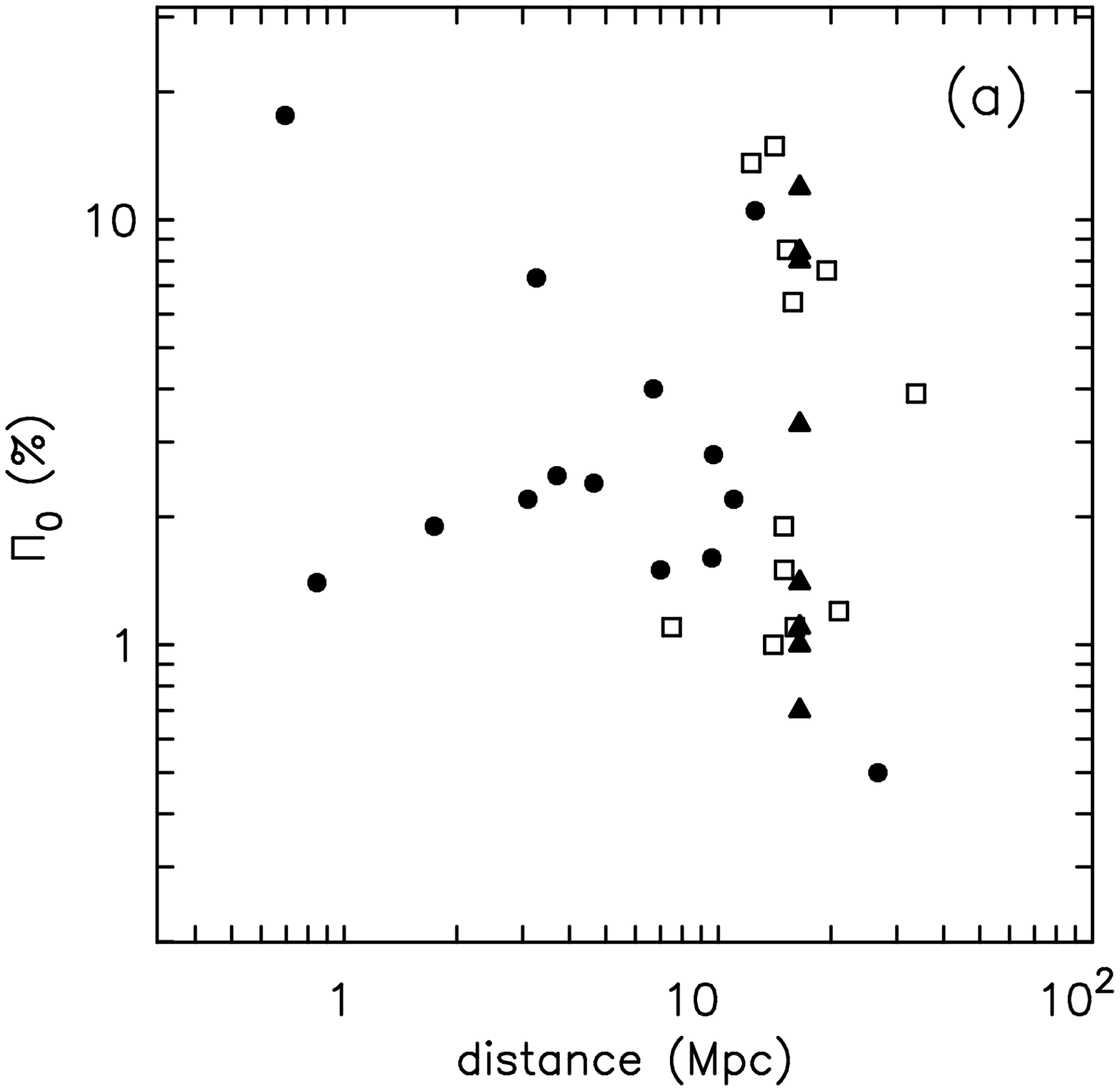}}
\resizebox{8.5cm}{!}{\includegraphics[angle=0]{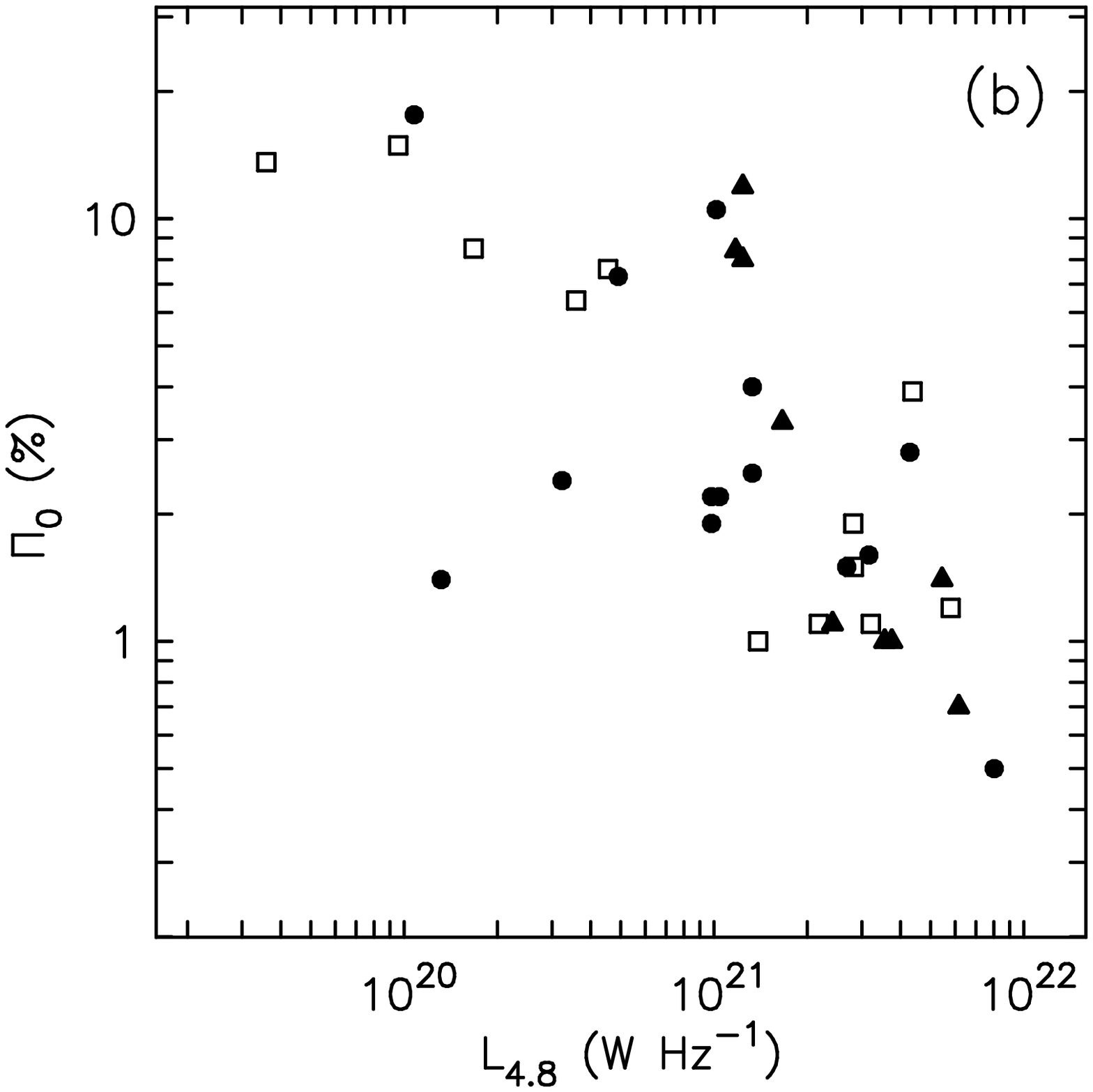}}
\caption{ Integrated fractional polarization as a function of distance
(a) and monochromatic 4.8 GHz luminosity (b) for the entire
sample. The symbols are the same as in Figure~\ref{PA-fig}. The
symbols for the Virgo galaxies NGC~4321 and NGC~4501 overlap
completely in (a), and partially in (b).
\label{PI_L-fig}
}  
\end{figure*}

In order to simulate unresolved observations from our maps of
nearby galaxies we integrated the Stokes I, Q, and U images of these
galaxies separately in ellipses. Their axial ratio was defined by the
outer isophotes in the Stokes $I$ image and the ellipses include the
entire galaxy up to the noise level of the Stokes I map. The
integrated $Q$ and $U$ flux densities were used to calculate the
integrated polarized flux density $S_p=\sqrt{(Q^2+U^2)}$, that was
corrected for polarization bias following \citet{simmons1985} to
obtain the intrinsic polarized flux density $S_{p,0}$. The correction
for polarization bias did not have a measurable effect for most
galaxies because of the high signal to noise ratio of the integrated
flux densities.  The integrated fractional polarization of each galaxy
was calculated as $\Pi_0 = S_{p,0}/S$, where $S$ is the total flux
density, including thermal and non-thermal emission. The polarization
angle $\theta_{\rm pol} = \onehalf \arctan(U/Q)$ of the integrated
emission is expected to be aligned with the apparent minor axis of the
optical galaxy.  The sensitivity of the observations is included in
the errors for $\Pi_0$ through the noise in the maps. The error
estimates also include the uncertainty in the zero level, that
dominates the uncertainty for galaxies with a large angular size such
as M~31. It is customary to list the position angle of the optical
major axis, and also to rotate the observed E-vectors by $90\degr$ to
obtain B-vectors in the plane of the sky.  We therefore define
$\thetaB = \theta_{\rm pol}+90\degr$ for comparison with published
position angles of the optical major axis, $\thetaopt$.  The results
are listed in Tables~\ref{nearby-tab}, \ref{virgo-tab}, and
\ref{barred-tab}.

Bandwidth depolarization is negligible in these data. For a
rotation measure as high as 300 rad m$^{-2}$, the polarized intensity
is decreased by less than 1\%. The integrated polarized flux densities
are not affected by significant residual instrumental
polarization. Small values for the integrated fractional polarization
are the result of integrating significant flux in Stokes Q and U, with
nearly equal positive and negative contributions. The uncertainty in
the polarization angle of the integrated emission is typically less
than 5 degrees.

Some galaxies (M~31, M~81, NGC~4565 in Table~\ref{nearby-tab},
NGC~4192, NGC~4302, NGC~4535 in Table~\ref{virgo-tab}, NGC~1300,
NGC~1493, NGC~3953, NGC~3992 in Table~\ref{barred-tab}) have
integrated fractional polarization $\Pi_0 > 7\%$.  Five of these
galaxies (M~31, NGC~4565, NGC~4192, NGC~1493, and NGC~3992) are more
than $10\%$ polarized after integration over solid angle.

Five of the Virgo galaxies have additional observations at 8.4 GHz.
The fractional polarization at both frequencies agrees to within a few
percent for these galaxies. Small differences can be attributed to
statistical errors, mainly at 8.4 GHz where the signal to noise ratio
of the emission is lower. The differences in $\Pi_0$ at 4.8 GHz and at
8.4 GHz per galaxy are much smaller than the differences between
galaxies. This confirms that the variation in $\Pi_0$ between galaxies
represents intrinsic differences between the galaxies. The
polarization angles at 4.8 GHz and 8.4 GHz are also mutually
consistent. In low-inclination galaxies, the polarization angle of the
integrated emission is more affected by asymmetries because the effect
of an axially symmetric component on the polarization vanishes mostly.
This can explain the difference in polarization angle for NGC~4303 ($i
= 18\degr$). The position angle of the major axis can also be affected
by ellipticity of the disk.

The fractional polarization of any particular galaxy depends on the
direction from which the galaxy is observed, as the projection of the
magnetic field on the plane of the sky and the path length of the line
of sight through the disk change. Any particular galaxy can only be
observed from one direction, but the variation of integrated
polarization properties with inclination can be revealed for a
sufficiently large sample of galaxies. The wide range in inclination
($20\degr < i < 90\degr$) in the sample allows an effective analysis
of the integrated polarization properties of spiral galaxies as a
function of inclination. It should be noted that the sample is not
randomly distributed in inclination.  The galaxies in
Tables~\ref{nearby-tab}, \ref{virgo-tab}, and \ref{barred-tab} were
observed to study their magnetic field configuration through imaging
polarimetry.  The sample is biased towards low inclination (face-on)
and high-inclination (edge-on) galaxies because these were preferred
for imaging of the magnetic field. In particular, the sub sample of
barred spirals in Table~\ref{barred-tab} contains only galaxies with
inclination $i < 60\degr$, with the exception of the added nearby
galaxy NGC~4631.  Any sample of spiral galaxies with inclination
derived from the apparent axial ratio appears deficient in galaxies
with $i \lesssim 20\degr$ because the disks of most spiral galaxies
are not exactly circular
\citep{binney1981,grosbol1985,lambas1992}. This effect is also
apparent in the present sample, but it does not affect our analysis.

Figure~\ref{diskpol_model-fig} shows $\Pi_0$ as a function of
inclination. The curves in Figure~\ref{diskpol_model-fig} represent
models that will be discussed in Section~\ref{model-sec}.  The formal
errors on $\Pi_0$ are generally small because of the high signal to
noise ratio in the integrated flux densities.  Some galaxies with
large angular size have larger errors because of uncertainties in the
zero level of the maps.  The large spread in $\Pi_0$, $1\% \lesssim
\Pi_0 \lesssim 15\%$, seen for each subsample reflects intrinsic
differences. The highest values of $\Pi_0$ in
Figure~\ref{diskpol_model-fig}a are found in the inclination range
from $\sim 50\degr$ to $\sim 80\degr$.  Some low-inclination galaxies
with high $\Pi_0$ are found among the barred galaxies in
Figure~\ref{diskpol_model-fig}b. The model curves in
Figure~\ref{diskpol_model-fig}b are shown for reference only, as they
do not include a bar.

Figure~\ref{PA-fig}a shows the correlation between the position angle
of the major axis, $\thetaopt$, and $\thetaB$ for galaxies in
Table~\ref{nearby-tab} and Table~\ref{virgo-tab}. If polarization of
the integrated emission is the result of projection of the azimuthal
magnetic field in the disk on the plane of the sky, a correlation is
expected between $\thetaB$ and the line of nodes of the disk
intersecting the plane of the sky. The latter is determined by the
position angle of the major axis, $\thetaB$.  The most deviating
galaxies in Figure~\ref{PA-fig}a are NGC~6946 and NGC~4254, with
inclination $i = 38\degr$ and $i = 42\degr$ respectively. Large-scale
departures from axial symmetry are most likely to be noticed for
galaxies with a small inclination. The position angles used in
Figure~\ref{PA-fig} are also less well defined.  At lower
inclinations, the position angle of the major axis is more difficult
to measure because the isophotes are nearly circular.  Besides
measurement uncertainties, the position angle of the major axis is not
a good estimator for the position angle of the line of nodes of the
disk if the inclination approaches $26\degr$, the inclination for
which a circular disk would have an apparent axial ratio of 0.9,
similar to the mean intrinsic ellipticity of spiral galaxy disks found by
\citet{lambas1992}. Consequently, the uncertainty associated
$\thetaopt$ is substantial but hard to quantify for galaxies with
inclination less than $\sim 40\degr$. For galaxies with higher
inclinations, $\thetaopt$ is estimated to be accurate to within $\sim
5\degr$. The uncertainty in $\thetaB$ is typically a few degrees, but
some additional scatter is expected because of differences in
structure between galaxies. The total scatter in Figure~\ref{PA-fig}a
for galaxies with inclination $\gtrsim 60\degr$ suggests that the
scatter introduced by structural differences is limited to $\pm
15\degr$.

Figure~\ref{PA-fig}b shows the same relation for the detected barred
galaxies in Table~\ref{barred-tab}. The scatter between $\thetaopt$
and $\thetaB$ appears substantially larger than in
Figure~\ref{PA-fig}a.  This may be a result of magnetic fields aligned
with the bar, where the orientation of the bar adds a parameter to the
problem. In this paper we focus mainly on galaxies without (strong)
bars, while noting that unresolved barred galaxies can in fact also be
polarized radio sources.

Figure~\ref{PI_L-fig} shows the integrated fractional polarization
$\Pi_0$ as a function of distance, and monochromatic luminosity
$L_{4.8}$. There should be no correlation between integrated
polarization and distance, and none is found in the present sample.
Figure~\ref{PI_L-fig}b shows a correlation between $\Pi_0$ and
$L_{4.8}$.  There are no galaxies in the sample with $L_{4.8}> 2
\times 10^{21}$ W Hz$^{-1}$ that have $\Pi_0 > 4 \%$.  Most galaxies
with $L_{4.8}< 5 \times 10^{20}$ W Hz$^{-1}$ do have $\Pi_0 > 4
\%$. As shown in Figure~\ref{diskpol_model-fig}, a galaxy can have low
$\Pi_0$ because of strong Faraday depolarization, or because of its
inclination. There is no correlation between inclination and
luminosity in the sample. The lower integrated polarization of
luminous galaxies in Figure~\ref{PI_L-fig}b suggests stronger Faraday
depolarization in these galaxies, or a lower degree of uniformity of
the magnetic field as will be discussed in Section~\ref{discussion1}.

\section{Models of the integrated polarization}
\label{model-sec}

\subsection{Polarized intensity in the disk}

The principal cause of net polarization of unresolved spiral galaxies
is the regular azimuthal magnetic field in the disk projected on the
sky along the apparent major axis. Faraday rotation and depolarization
are affected by the path length through the disk, the line-of-sight
component of the regular magnetic field, and fluctuations in thermal
electron density and magnetic field strength.  A first analysis of the
integrated polarization of spiral galaxies is best served with the
simplest possible model that includes these factors.  We consider
axially symmetric models with a regular azimuthal magnetic field in
the disk, and an isotropic random component of the magnetic field.

Using this simple geometry, the integrated polarization of a spiral
galaxy is modeled by considering each solid angle element of the disk
as a slab with a specified line-of-sight component of the
regular magnetic field that depends on inclination and location in the
disk. \citet{sokoloff1998} provided solutions of polarized intensity
from a slab of Faraday-rotating medium mixed with relativistic
electrons that produce synchrotron emission. The polarized intensity
is expressed as a complex quantity $\mathcal{P} = Q + j U$
\citep{spangler1982}, where $Q$ and $U$ are Stokes parameters that
describe the linear polarization. Following the conventions adopted by
\citet{sokoloff1998}, $\mathcal{P}$, $Q$, and $U$ are normalized by
the total intensity of the synchrotron radiation, so that the modulus
of $\mathcal{P}$ is the fractional polarization. The polarization
angle of the complex polarization is $\Psi = \onehalf \arctan (U/Q)$.

At higher frequencies, a contribution from thermal emission to the
total intensity reduces the observed fractional polarization by a
factor ($1-f_{th}$), where $f_{th}$ is the fraction of the total
intensity generated by thermal emission. We include thermal emission
by setting $f_{th} = 0.23$ at 4.8 GHz
\citep{condon1990,condon1992,niklas1997}.  Determination of $f_{th}$
in individual galaxies is difficult because of the need to separate
thermal and non-thermal emission in the radio spectrum.  The rms
scatter in $f_{th}$ from galaxy to galaxy is believed to be at most a
factor of 2 \citep{condon1992}. This translates into a spread of $\sim
25\%$ in ($1-f_{th}$) from the adopted value of 0.77.

\begin{deluxetable}{lrcccccccc}
\tablecolumns{10}
\tablewidth{0pc} 
\tabletypesize{\scriptsize}
\tablecaption{ Sensitivity to model parameters$^a$}
\tablehead{
\colhead{Inclination} & \colhead{$B$} & \colhead{$f_{\rm B}$}  & \colhead{$n_e$} & \colhead{$f_{i}$} & \colhead{$l_{\rm turb}$} & \colhead{$2h$}
} \startdata $i = 40\degr$ & 0.007 & 1.01 & 0.006 & 0.034 & 0.042 &
0.035 \\ $i = 60\degr$ & 0.17 & 1.09 & 0.17 & 0.12 & 0.057 & 0.006 \\
$i = 75\degr$ & 0.50 & 0.83 & 0.51 & 0.45 & 0.055 & 0.45 \\ \enddata
\tablenotetext{a}{Numbers listed are normalized contributions to the
total scatter in a Monte Carlo simulation with 10000 realizations per
inclination per parameter, where all parameters were drawn independently from a
uniform distribution. Parameters with a small contribution to the
total scatter are relatively unimportant. Parameter ranges adopted in
this simulation: $1 < B < 10$ $\mu$G, $0.7 < f_{\rm B} < 4.0$, $0.01 <
n_e < 0.1$ $\rm cm^{-3}$, $0.1 < f_i < 1.0$, $10 < l_{turb} < 100$ pc,
$200 < 2h < 2000$ pc.}
\label{model_var-tab}
\end{deluxetable}

In this paper we adopt a solution for the polarized intensity from a
uniform slab of finite thickness along the line of sight from
\citet{sokoloff1998}. For a given inclination $i$, the component of
the regular magnetic field along the line of sight $B_\|$ and
perpendicular to the line of sight $B_\bot$ can be evaluated. The
uniformity of the magnetic field is expressed as the ratio of the
isotropic random magnetic field strength to the regular magnetic field
strength, defined as $f_B=\sigma_{\rm B}\sqrt{3}/B$, where
$\sigma_{\rm B}$ is the rms strength of the random component of the
magnetic field in one direction.

The complex polarization can be written as the product of
a wavelength-independent part that describes the polarization if
Faraday rotation effects can be neglected, and a wavelength-dependent
part that includes Faraday rotation effects. The
wavelength-independent part of the complex polarization is
\begin{equation}
\langle {\mathcal P_0} \rangle = p_i {B_\bot^2 \over B_\bot^2 + 2\sigma_{\rm B}^2 } \exp\bigl [ 2j ( \onehalf \pi + \arctan(B_y/B_x) ) \bigr ],
\label{P0-eq}
\end{equation}
where $p_i \approx 0.75$ is the intrinsic maximum polarization for
synchrotron emission in a regular magnetic field, and $B_x$ and $B_y$
are the components of the regular magnetic field along two orthogonal
axes in the plane of the sky. Equation~\ref{P0-eq} assumes a constant
spectral index of $-1$ for the synchrotron emission of a galaxy.  We
define the $x-axis$ along the apparent major axis of the disk. The
brackets $\langle \cdot \rangle$ indicate that the emission is
averaged over a volume in the disk, that is sufficiently narrow to keep
the regular magnetic field constant inside the volume, but extends
along the line of sight through the disk.  This is a reasonable
approximation except for very high inclinations $i \gtrsim 80\degr$,
where the azimuthal field must change along the line of sight, but
only if the line of sight is near the major axis.

Equation~\ref{P0-eq} includes the effect of decreased polarization
because of the presence of a random component of the magnetic field in
the plane of the sky. \citet{sokoloff1998} referred to this as
wavelength-independent depolarization. The term beam depolarization is
sometimes used in this situation, but this term is used also in case
of wavelength-dependent Faraday effects.  The plane of polarization is
perpendicular to the component of the regular magnetic field projected
on the plane of the sky.

\begin{figure}
\resizebox{\columnwidth}{!}{\includegraphics[angle=0]{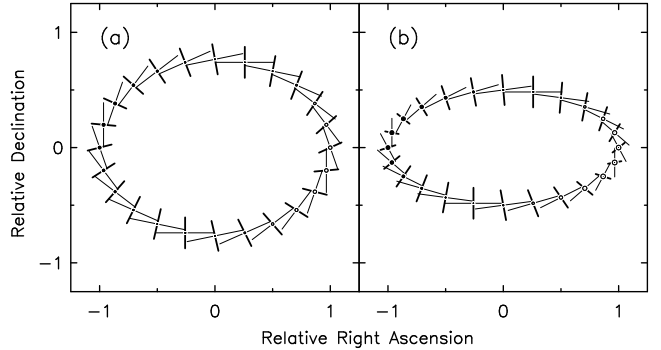}}
\caption{ Representation of polarized intensity in an annulus of a
spiral galaxy disk for $i=40\degr$ (a), and $i=60\degr$ (b). Thin
lines represent the component of the magnetic field perpendicular to
the line of sight, $B_\bot$.  The pitch angle of the field is
$15\degr$ in this example. Thick lines represent the polarization
E-vectors, with length proportional to the polarized intensity. The
line-of-sight component of the magnetic field $B_\|$ is indicated by
open circles (towards the observer) and filled circles (away from the
observer) with size proportional to $B_\|$. The integrated fractional
polarization in (a) is $\Pi_0 = 6.1\%$, and in (b) $\Pi_0 = 13.1\%$,
including a thermal fraction $f_{th}=0.23$ at 4.8 GHz. The separation
of the positions shown in this figure does not represent the step size
of the integration.
\label{spiral_proj-fig}
}  
\end{figure}

The observed polarization $\langle {\mathcal P} \rangle$ differs from
$\langle {\mathcal P_0} \rangle$ because the polarized synchrotron
emission generated in a small volume element in the disk is Faraday
rotated as it travels through the disk to the observer. The effects of
Faraday rotation can be written as a wavelength-dependent factor in
the complex polarization, that includes an integral along the line of
sight.  \citet{burn1966} and \citet{sokoloff1998} evaluated this
integral analytically for a uniform slab.  The observed polarized
intensity from a small part of the disk is
\begin{equation}
\langle{\mathcal P} \rangle = \langle {\mathcal P_0} \rangle {1 - \exp(-S) \over S},
\label{P-eq}
\end{equation}
where $S = 2 \lambda^4 \sigma_{\rm RM}^2 -2 j \lambda^2 {\mathcal R}$
is a complex number that includes rotation measure fluctuations
$\sigma_{\rm RM}$ in the real part and the Faraday depth $\mathcal R$
in the imaginary part.

The real part of $S$ contains fluctuations in rotation measure along
the line of sight through the factor $\sigma_{\rm RM}$, which
represents the rms amplitude of rotation measure fluctuations after
integration along the line of sight. The plane of polarization makes a
random walk because of fluctuations in density and magnetic
field. When polarized emission is integrated over different lines of
sight, or different depths inside the source, depolarization of the
emission results.  This process is called internal Faraday dispersion.
Evaluation of $\sigma_{\rm RM}$ requires assumptions about the
correlation scale of fluctuations in density and magnetic field
strength. \citet{sokoloff1998} discuss the difficulties estimating a
correlation scale length, and assume that the scale length is the same
for fluctuations in density, magnetic field, and rotation measure. The
strength of the magnetic field in a turbulent cell defined by the
correlation length is $\sigma_{\rm B}$.  Rotation measure fluctuations
also depend on the path length $L$ through the disk with thickness $2
h$, observed at inclination $i$, given by $L = 2h / \cos i$.  The
line-of-sight filling factor of the Faraday-rotating medium is $f_i$.
Rotation measure fluctuations are then calculated following
\citet{sokoloff1998}
\begin{equation}
\sigma_{\rm RM} = 0.81 n_e \sigma_{\rm B} l_{\rm turb} \Bigl({ 2h f_i \over l_{\rm turb} \cos i } \Bigr)^{\onehalf}
\end{equation}

The imaginary part of $S$ describes rotation of the plane of
polarization by the line-of-sight component of the regular magnetic
field. The Faraday depth $\mathcal{R} = n_e f_i B_\| 2h/\cos i $ is
the product of the mean electron density ($n_e f_i$), where $n_e$ is
the electron density in a turbulent cell, $B_\|$ the line-of-sight
component of the regular magnetic field, and $h$ the path length
through the disk. Rotation of the plane of polarization also
depolarizes emission integrated along the line of sight as emission
from different depths of the source is rotated by different amounts.
This mechanism for depolarization is called differential Faraday
rotation.  It is sometimes called depth depolarization, but it should
be understood that there is a different depolarizing effect in the
form of Faraday dispersion that is also the result of integrating
emission along the line of sight.

\subsection{Integrated polarized emission}

\begin{deluxetable*}{lrrccccccc}
\tablecolumns{10}
\tablewidth{0pc} 
\tabletypesize{\scriptsize}
\tablecaption{ Model parameters for Figure~\ref{diskpol_model-fig}}
\tablehead{
\colhead{Model} & \colhead{$B$} & \colhead{$f_{\rm B}$}  & \colhead{$n_e$} & \colhead{$f_{i}$} & \colhead{$l_{\rm turb}$} & \colhead{$2h$} & \colhead{$f_{th}$
} & \colhead{$\Pi_{\rm max}$} & \colhead{$\Pi_{\rm med}$}  \\
\colhead{\ } & \colhead{($\rm \mu G$)} & \colhead{\ }  & \colhead{($\rm cm^{-3}$)} & \colhead{\ } & \colhead{(pc)} & \colhead{(pc)} & \colhead{\ }  & \colhead
{(\%)} & \colhead{(\%)}
}
\startdata
1$^{a}$      & 5.0  & 1.0   &  0.03   &  0.5  & 50  &  1000  & 0.23  & 16.4 & 10.2 \\
2$^{b}$      & 10.0  & 1.0   &  0.03   &  0.5  & 50  &  1000 & 0.23  & 13.7 &  8.2 \\
3$^{c}$      & 5.0  & 1.0   &  0.09   &  0.5  & 50  &  1000  & 0.23  & 12.0 &  7.0 \\ \\
4$^{a}$      & 5.0  & 2.0   &  0.03   &  0.5  & 50  &  1000  & 0.23  & 6.8  &  4.7 \\ 
5$^{b}$      & 10.0  & 2.0   &  0.03   &  0.5  & 50  &  1000 & 0.23  & 5.6  &  3.7 \\
6$^{c}$      & 5.0  & 2.0   &  0.09   &  0.5  & 50  &  1000  & 0.23  & 4.8  &  3.1 \\ \\
7$^{a}$      & 5.0  & 3.0   &  0.03   &  0.5  & 50  &  1000  & 0.23  & 3.4  &  2.4 \\
8$^{b}$      & 10.0  & 3.0   &  0.03   &  0.5  & 50  &  1000 & 0.23  & 2.7  &  1.8 \\
9$^{c}$      & 5.0  & 3.0   &  0.09   &  0.5  & 50  &  1000  & 0.23  & 2.2  &  1.4 \\
\enddata
\tablenotetext{a}{Models indicated by a solid curve in Figures~\ref{diskpol_model-fig} and \ref{Pi_PDF-fig}}
\tablenotetext{b}{Models indicated by a dashed curve in Figures~\ref{diskpol_model-fig} and \ref{Pi_PDF-fig}}
\tablenotetext{c}{Models indicated by a dotted curve in Figures~\ref{diskpol_model-fig} and \ref{Pi_PDF-fig}}
\label{model_par-tab}
\end{deluxetable*}

Equation~\ref{P-eq} provides the complex polarized intensity
$\langle{\mathcal P} \rangle$ for every line of sight, from which we
obtain the local Stokes $Q$ and $U$ intensities, normalized by the
local total intensity that is presumed uniform in the current
models. The model Stokes $Q$ and $U$ intensities can be integrated
over the disk to predict the model polarized flux density, fractional
polarization $\Pi_0$, and the angle $\thetaB$ of an unresolved spiral
galaxy. The integration over the disk reduces to a one-dimensional
numerical integration over azimuthal angle for the symmetric models
under consideration.

Assuming values for the various model parameters, the integration over
the disk can be performed as a function of inclination.  The
integrated polarization is more sensitive to certain model parameters,
and the influence of model parameters can change with inclination. The
relative importance of the parameters in our model was investigated
through a Monte Carlo experiment. Each parameter was varied
separately, and the resulting range of the integrated polarization was
divided by the range obtained when all parameters were varied
independently at the same time. The spread in integrated fractional
polarization was taken as the difference between the upper quartile
and the lower quartile. As the underlying distributions are strongly
skewed, the quartiles provide a better indication of the spread in the
distribution than the standard deviation. The normalization with the
``total'' scatter allows us to compare the importance of parameters
between the three inclinations. It should be noted that the non-linear
character of the model, and the opposing effects of certain parameters
can make the ratio of the normalized quartile ranges larger than
unity.

Table~\ref{model_var-tab} shows for three values of the inclination
the result of 10000 realizations of each parameter, drawn from a
uniform distribution. The magnetic field strength, electron density,
filling factor, turbulent length scale, and the vertical thickness of
the disk were varied by an order of magnitude, as indicated in the
footnote to Table~\ref{model_var-tab}. The ratio of random to regular
magnetic field strength ($f_{\rm B}$) was varied over a smaller range
(0.7 to 4.0). At low inclinations, the normalized quartile range for
variation of $f_{\rm B}$ is 1.01, while the normalized quartile ranges
of all other parameters are less than 0.1. Clearly, the ratio of
random to regular magnetic field dominates the fractional polarization
of the model for small inclinations. At $i=60\degr$, $f_{\rm B}$ is
still the most important parameter, but other parameters such as the
regular magnetic field strength and the electron density become more
important. At $i=75\degr$, $f_{\rm B}$ is still important, but the
parameters that define $\mathcal{R} = n_e f_i B_\| 2h/\cos i$ are
almost equally important. The relative importance of $n_e$, $f_i$,
$B$, and $2h$ is similar at $i=75\degr$, because they contribute
equally to $\mathcal R$, although subtle differences occur because of
the dependence on $\sigma_{\rm RM}$. The turbulent scale length
$l_{turb}$ is the parameter in our model that is least constrained by
observations.  Table~\ref{model_var-tab} shows that its effect on our
results is insubstantial.

The simplest model is a disk with uniform parameters throughout that
can be represented by a set of concentric annuli, all with the same
integrated polarization properties. The model makes no assumptions
about the magnetic field structure in the center of the disk. This is
not a significant simplification for most spiral galaxies, where most
of the polarized emission originates in the disk.
Figure~\ref{spiral_proj-fig} shows a graphic representation of
polarized intensity at a representative subset of locations (a ring)
in the disk for model 1 in Table~\ref{model_par-tab} at inclinations
$i=40\degr$ and $i=60\degr$. Thin lines indicate the magnitude and
direction of the magnetic field component in the plane of the sky,
$B_\bot$. Thick lines represent polarization vectors. Open and filled
circles represent the component of the regular magnetic field along
the line of sight, $B_\|$.

At low inclination (Figure~\ref{spiral_proj-fig}a), the path length
through the disk is relatively small and the line-of-sight component
of the regular field is also small.  The polarized intensity in
Figure~\ref{spiral_proj-fig}a is fairly uniform across the disk and
relatively high. At higher inclination, the path length through the
disk increases, and the line-of-sight component of the regular
magnetic field increases, in particular near the major axis. The
polarized intensity is smaller everywhere in the disk, but most
strongly near the major axis. Here, the line-of-sight component of the
azimuthal regular field is largest, so Faraday rotation is strongest
and the component $B_\bot$ (Equation~\ref{P0-eq}) is smallest.
Faraday depolarization is therefore strongest near the major axis, and
the intrinsic polarization of the emission is smaller than elsewhere
in the disk.  Faraday rotation of the plane of polarization is also
stronger near the major axis in Figure~\ref{spiral_proj-fig}.

The model Stokes $Q$ and $U$ intensities, calculated by solving
Equation~\ref{P-eq} for every line of sight through the disk, are
integrated over the disk to predict the polarized intensity of an
unresolved spiral galaxy. The integration over the disk reduces to a
one-dimensional numerical integration over azimuthal angle for the
symmetric models under consideration.  The model in
Figure~\ref{spiral_proj-fig}a has integrated fractional polarization
$\Pi_0 = 7.9\%$, while the model in Figure~\ref{spiral_proj-fig}b has
integrated fractional polarization $\Pi_0 = 17.0\%$. Despite the
stronger depolarization anywhere in the disk at inclination $i =
60\degr$, the {\it integrated } polarization is stronger at $i =
60\degr$ than at $i = 40\degr$, because of the smaller degree of
symmetry of the polarized emission.

The model predictions for $\Pi_0$ and $\thetaB$ are independent of
the spiral pitch angle of the magnetic field. For a disk with constant
thickness and uniform properties, the path length through the disk
depends only on inclination, and the polarized intensity depends only
on the angle of the regular component of the magnetic field with the
line of sight. The effect of a pitch angle of the magnetic field, is
to change the azimuthal angle in the disk where the line of sight
makes a certain angle with the regular magnetic field. The integral
over all azimuthal angles in the disk is not affected by a translation
in azimuthal angle resulting from the pitch angle of the magnetic
field.

Table~\ref{model_par-tab} gives values for the parameters in the
models that are compared with the data. Model 7 corresponds with the
values adopted by \citet{beck2007} for NGC~6946. The filling factor of
ionized gas $f_{i}$ and the thickness of the disk $2h$ were kept
constant, because their effect on the models is similar to that of the
electron density and the magnetic field.  Model curves of fractional
polarization as a function of inclination are shown in
Figure~\ref{diskpol_model-fig}. Three families of models are shown
(Table~\ref{model_par-tab}), with $f_{\rm B} = $ 1, 2, and 3. For each
value of $f_{\rm B}$, model curves are shown for $n_e = 0.03$
cm$^{-3}$, $B = 5\ \mu$G (solid curves), $n_e = 0.03$ cm$^{-3}$, $B =
10 \mu$G (dashed curves), and $n_e = 0.09$ cm$^{-3}$, $B = 5\ \mu$G
(dotted curves).

At low inclination, the models are mainly distinguished by the value
of $f_{\rm B}$. The line-of-sight component of the regular magnetic
field is small at low inclination, and the main depolarization
mechanisms are wavelength independent depolarization and Faraday
dispersion. At inclination $i \gtrsim 50\degr$, differential Faraday
rotation along the line of sight becomes increasingly important, and
significant effects of the strength of the regular magnetic field and
the density of thermal electrons are found.

This effect already occurs at low inclination, where the line-of-sight
component of the magnetic field near the major axis increases
proportionally to $i$, but the path length through the disk $\sim
1/\cos i$ is constant to first order in $i$. The result is that models
that include Faraday depolarization predict a {\it higher} degree of
polarization at low inclinations than models that do not include
Faraday depolarization \citep{stil2007a}. The present models also
predict a higher fractional polarization at low inclination if the
electron density $n_e$ or the strength of the regular magnetic field
is increased, if the ratio of random magnetic field to regular
magnetic field strength is not too high.

The models more or less cover the same area of
Figure~\ref{diskpol_model-fig}a as the data. This result supports the
idea that the observed integrated polarization of spiral galaxies is
related to the large-scale field in the disk.  The most notable
difference between the data and the models is the observed low $\Pi_0
\lesssim 2\%$ of some galaxies with intermediate inclination, $40\degr
\lesssim i \lesssim 60\degr$. It is difficult to change any single
parameter in the model to obtain a fractional polarization $\Pi
\lesssim 2\%$ in this inclination range, unless higher values of
$f_{\rm B}$ are assumed.

\subsection{Probability distribution of $\Pi_0$}
\label{Pi_PDF-sec}

The results in the previous sections suggest that the integrated 4.8
GHz emission of spiral galaxies can show substantial
polarization. Deep polarization surveys can detect distant unresolved
spiral galaxies. How many spiral galaxies can be detected, and the
interpretation of unresolved galaxies detected in polarization, is
determined by the probability density function (PDF) of the fractional
polarization of spiral galaxies. The wavelength dependence of
integrated polarization is also best summarized by means of the PDF of
fractional polarization.  In this section we discuss the expected
shape of the PDF of fractional polarization for unresolved spiral
galaxies at 4.8 GHz and 1.4 GHz. We also discuss the PDF of $\Pi_0$ in
the special case of low-inclination galaxies ($i \lesssim 30\degr$),
where the integrated polarization depends mostly on Faraday
dispersion, but the actual value of the inclination cannot be
determined reliably from the optical axial ratio.
 
\begin{figure}
\resizebox{\columnwidth}{!}{\includegraphics[angle=0]{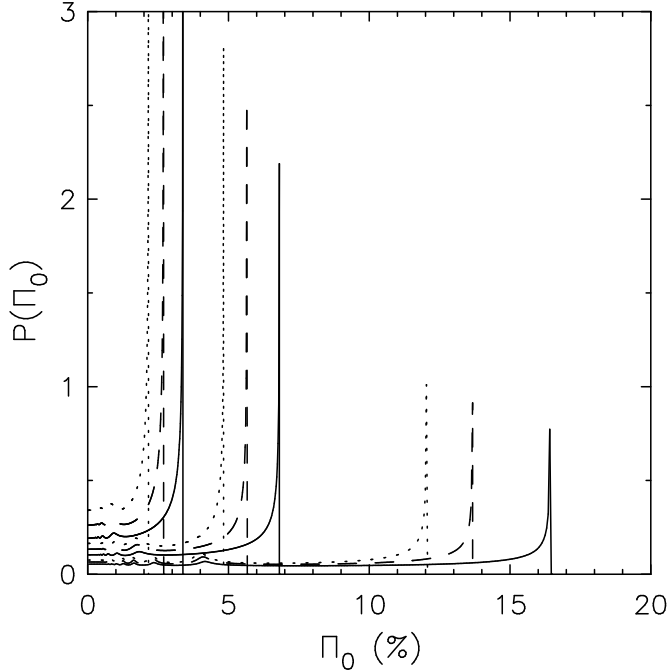}}
\caption{ Probability density functions (PDF) of $\Pi_0$ for
randomly-oriented spiral galaxies derived from the models in
Figure~\ref{diskpol_model-fig}. The continuous, dashed, and dotted
curves correspond with models 1 -- 9 in Table~\ref{model_par-tab}, in
order of decreasing maximum $\Pi_{\rm max}$. The curves are normalized
such that the integral of the PDF is 1.  Median values $\Pi_{\rm med}$
of these distributions are listed in Table~\ref{model_par-tab}.
\label{Pi_PDF-fig}
}  
\end{figure}

\begin{figure*}
\resizebox{\textwidth}{!}{\includegraphics[angle=-90]{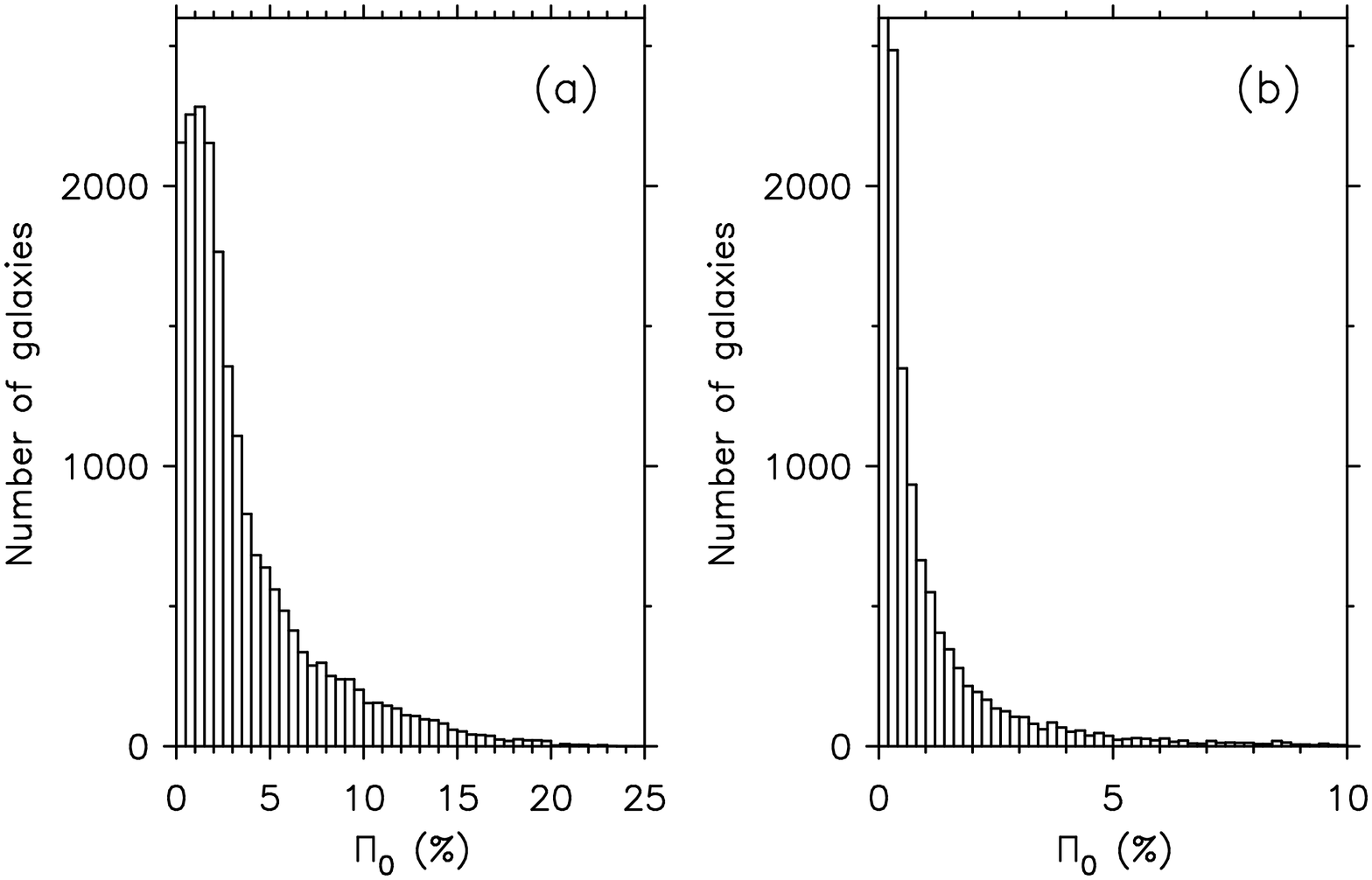}}
\caption{ (a) Simulated histogram of fractional polarization of 20,000
randomly oriented model spiral galaxies at 4.8 GHz. Ranges of model
parameters used for this simulation are given in
Table~\ref{model_var-tab}.  The fractional polarization for the same
galaxies at 1.4 GHz is shown in (b).  The vertical axis in (b) is
truncated (lowest bin only).
\label{Pi_PDF-sim-fig}
}  
\end{figure*}

The models for the integrated fractional polarization as a function of
inclination can be used to derive a probability density function (PDF)
for $\Pi_0$ of unresolved, randomly oriented spiral galaxies. This PDF
can be used to predict the number of spiral galaxies detectable in
deep polarization surveys based on population models of the total
intensity source counts \citep{stil2007a}. The shape of the PDF can be
modeled as a function of rest-frame wavelength and fitted to
observations. \citet{taylor2007} used Monte-Carlo simulations of noise
statistics and observational selection effects, and derived the shape
of the PDF for $\Pi_0$ for faint polarized sources from a maximum
likelihood fit to the data. A similar approach can be used in the
future to compare the PDF of $\Pi_0$ for galaxies at high redshift
with the PDF for local galaxies.

The PDF for $\Pi_0$ follows from the probability $P(i)$ to observe a galaxy at
inclination between $i$ and $i+di$,
\begin{equation}
P(i)di = \sin(i) di \ \ \ \ \  (0\degr \le i \le 90\degr).
\end{equation}
The probability to find a fractional polarization between $\Pi_0$ and
$\Pi_0+\Delta\Pi_0$ is equal to the probability to find a galaxy with
inclination in one of up to two inclination ranges $i_{{\rm min}}$ to
$i_{{\rm max}}$ that yield the same fractional polarization. Therefore,
\begin{equation}
P(\Pi_0) \Delta\Pi_0 = \sum_{k = 1}^N ( \cos i_{{\rm min},k} - \cos i_{{\rm max},k}) 
\label{PIdist-eq}
\end{equation}
In most cases, $N=2$, except at the maximum of $\Pi_0$, where
$N=1$.

The PDFs of $\Pi_0$ for models 1 -- 9 in Table~\ref{model_par-tab} are
shown in Figure~\ref{Pi_PDF-fig}.  The median inclination for
randomly-oriented galaxies is $i_{\rm med}=60\degr$. Nearly half
(47\%) of the galaxies in a randomly oriented set have inclination
between $50\degr$ and $80\degr$, for which $\Pi_0$ is near the maximum
of the curves in Figure~\ref{diskpol_model-fig}a.  Consequently, all
models predict a PDF that is sharply peaked toward the maximum
possible value of $\Pi_0$.

Qualitatively, the shapes of the PDFs in Figure~\ref{Pi_PDF-fig} do
not depend strongly on the details of the models. The PDFs are derived
from the curves in Figure~\ref{diskpol_model-fig}. The shapes of these
curves can be understood qualitatively in terms of two competing
effects.  If the inclination is not too large, the integrated
polarization increases with inclination because of a larger net
magnetic field in the plane of the sky.  At higher inclinations,
Faraday depolarization dominates and the integrated polarization
decreases. A broad maximum in $\Pi_0$ must occur at intermediate
inclination, that gives rise to the peak of the PDF near the maximum
of $\Pi_0$.

Another feature of the PDFs in Figure~\ref{Pi_PDF-fig} is that they
converge to a constant value for small $\Pi_0$. The distribution is
dominated here by galaxies at low inclination, with a smaller
contribution of highly inclined galaxies with strong Faraday
depolarization.  Small fluctuations in the curves at small $\Pi_0$ are
the result of occurrences where the Faraday rotation of the disk at
the major axis is $\mathcal{R}\lambda^2 = n \pi$ ($n = 1, 2, 3,
\ldots$) at high inclination. These fluctuations are small, and would
not be noticed in observations.

The plateau in the distributions in Figure~\ref{Pi_PDF-fig} can be
understood based on a simple analytic argument.  For inclination $i
\lesssim 30\degr$, the model curves in Figure~\ref{diskpol_model-fig}
can be approximated with the relation $\Pi_0 = a i^2$, with the
inclination $i$ in radians and $a$ a constant that depends on the
model. The value of $a$ may depend on complicated physics, but the
only assumption here is that the increase of $\Pi_0$ for small $i$ can
approximated by a parabola, for fixed parameters of the model. The
probability to find a fractional polarization between $\Pi_0$ and
$\Pi_0+d\Pi_0$ is equal to the probability to find a galaxy with
inclination in the interval $i_1 \leqq i \leqq i_2 < 30\degr$, with
$\Pi_0 = a i_{{\rm min},1}^2$, and $(\Pi_0+d\Pi_0) = a i_{{\rm
max},1}^2$.  Equation~\ref{PIdist-eq} reduces to
\begin{equation}
P(\Pi_0)d\Pi_0 = \cos i_{{\rm min},1} - \cos i_{{\rm max},1} = {1 \over 2a} \Bigl( {\sin i \over i}\Bigr) d\Pi_0.
\label{low_i-eq}
\end{equation}
The last step used that $(i_{{\rm max},1}-i_{{\rm min},1})$ is the
infinitesimal increment $di = d\Pi_0/(2 a i)$, and that
$\onehalf(i_{{\rm min},1}+i_{{\rm max},1})=i$.  For $i < 30\degr$ the
factor $\sin i / i \approx 1$ to an accuracy better than 5\%, so
$P(\Pi_0)$ is a constant that depends only on $a$.  The models
presented in Section~\ref{model-sec} suggest that $a$ depends mainly
on the ratio of the random magnetic field to the regular magnetic
field. This is why models 1, 2, and 3 converge to the same value for
small $\Pi_0$, and models 4, 5, and 6 converge to a larger value. The
differences between models 7, 8, and 9 arise from the more significant
contributions from galaxies with a high inclination. These models also
converge to a common value if only galaxies with low inclination are
considered.

The distribution of $\Pi_0$ of an observed sample of spiral galaxies
would be a superposition of a continuous set of model distributions
similar to the nine curves in Figure~\ref{Pi_PDF-fig}. A simulated PDF
for 20,000 galaxies with model parameters varied over the ranges given
in Table~\ref{model_var-tab} is shown in Figure~\ref{Pi_PDF-sim-fig}.
At 4.8 GHz (Figure~\ref{Pi_PDF-sim-fig}a), the distribution of $\Pi_0$
is quite broad, with a peak between $1\%$ and $2\%$ polarization.  The
location of the peak and the depth of the minimum near $\Pi_0 = 0\%$
depends on the number of galaxies with strong Faraday depolarization.
At 1.4 GHz (Figure~\ref{Pi_PDF-sim-fig}b), Faraday depolarization is
much stronger and the distribution is sharply peaked towards $\Pi_0 =
0\%$.  For the distributions in Figure~\ref{Pi_PDF-sim-fig} we find
that 78\% of galaxies has $\Pi_0 > 1\%$ at 4.8 GHz, while at 1.4
GHz 17\% has $\Pi_0 > 1\%$. The broad tail of the distribution in
Figure~\ref{model_var-tab}b contains galaxies with low to moderate
inclination. The median inclination for galaxies with $\Pi_0 > 5\%$ at
1.4 GHz is $41\degr$.

Integrated polarimetry of an unbiased sample of a few hundred spiral
galaxies would allow a statistical analysis for the $\Pi_0$
distributions of spiral galaxies, including subsets by inclination,
morphological type and star formation rate. The $\Pi_0$ distribution
can be retrieved from the data in the presence of noise through
maximum likelihood fits presented by \citet{taylor2007}.  The skewed
shape on the PDF in Figure~\ref{Pi_PDF-sim-fig}a, suggests that the
best theoretical function to represent the PDF of spiral galaxies
would be the lowest order {\it anti}-symmetric deviation from a
Gaussian that shifts the peak of the distribution to a finite value of
$\Pi_0$. In terms of the Gauss-Hermite series used in
\citet{taylor2007}, the lowest order term would include coefficient
$h_3$ \citep{vdmarel1993}.

\section{Discussion}
\label{discussion-sec}

\subsection{Diagnostics from integrated polarization}
\label{discussion1}

We presented observations of the integrated polarization of nearby
spiral galaxies to study the polarization properties of spiral
galaxies as unresolved radio sources. It is shown that unresolved
spiral galaxies are polarized radio sources with fractional
polarization up to $\sim 20\%$ at 4.8 GHz.  This high degree of
fractional polarization arises from the regular azimuthal component of
the magnetic field in the spiral galaxies observed, that is projected
into a component in the plane of the sky that is predominantly
oriented along the apparent major axis of the disk seen at inclination
$i$.  The projected magnetic field direction as implied by the plane
of polarization, rotated by $90\degr$, is well correlated with the
position angle of the major axis for galaxies observed at higher
inclination.  The highest integrated polarization is expected for
galaxies at intermediate inclination. For a sample of randomly
oriented spiral galaxies, 47\% has an inclination between $50\degr$
and $80\degr$.  Galaxies in this inclination range are
under-represented in the present sample because the data used in this
paper were obtained for polarization studies that preferred
low-inclination or high-inclination galaxies. The number of galaxies
with a high fractional polarization in Figure~\ref{diskpol_model-fig}
may therefore be under represented compared with an unbiased sample.

Three of the eight Virgo spirals in our data set have integrated
fractional polarization $\Pi_0 \geqq 8\%$. The range of $\Pi_0$ of the
Virgo spirals is similar to that of the nearby ``field'' spirals.  We
do not find evidence in the present data that the integrated
polarization of spiral galaxies in a cluster environment would be
different from field galaxies. Although cluster galaxies are distorted
by interactions with other galaxies and the intracluster medium
\citep{wezgowiec2007,chyzy2007,vollmer2007}, the Virgo galaxies in our
sample cannot be distinguished from field galaxies in their integrated
polarization. This result should be verified with a larger,
complete sample of cluster galaxies.

The barred galaxies in Table~\ref{barred-tab} show that a high
integrated polarization also occurs for barred galaxies. In some of
these barred galaxies, polarized emission from the bar is negligible
compared to polarized emission from the disk. In other barred spirals,
polarized emission from the bar region dominates, and the integrated
polarization depends on the orientation of the bar, as well as the
inclination of the galaxy. The effect of a bar is not included in the
present models, so the existence of significant departures from the
models is not unexpected. Non-axially symmetric models of the
integrated polarization of spiral galaxies are deferred to a later
paper.

Figure~\ref{PI_L-fig} suggests a difference between luminous and less
luminous spiral galaxies. The three sub samples separately also show a
smaller integrated fractional polarization for luminous spiral
galaxies.  Face-on galaxies and depolarized edge-on galaxies should
exist at every luminosity, although they form a minority in a set of
randomly oriented spiral galaxies. The presence of some galaxies with
low $\Pi_0$ at every luminosity in Figure~\ref{PI_L-fig} is therefore
not surprising.  The luminous galaxies in the sample cover the same
range in inclination as the galaxies with lower luminosity. This
excludes selection effects in inclination as a possible origin of the
correlation.  No selection effect could be identified that would
exclude luminous galaxies with a high $\Pi_0$ from the sample.

The lack of luminous galaxies with a high $\Pi_0$ gains significance
in view of the results from Section~\ref{Pi_PDF-sec}. Given certain
physical conditions, the inclination dependence of $\Pi_0$ favors high
values of $\Pi_0$. The lower polarization of luminous spiral galaxies
therefore indicates different conditions in these galaxies. A higher
rate of star formation increases the random component of the magnetic
field through the effect of supernova explosions and stellar winds. A
stronger total magnetic field and lower polarization are found locally
in regions of galaxies where the star formation rate is high
\citep{beck2005,chyzy2008}. The correlation in Figure~\ref{PI_L-fig}b
may be explained similarly in terms of a generally higher star
formation rate.  The correlation of polarization with luminosity is an
important factor for predictions of the number of spiral galaxies that
will be detected in deep polarization surveys.  Larger samples of
spiral galaxies are required to confirm this result, and to allow more
detailed modeling.

Simple axially symmetric models for the integrated polarization of
spiral galaxies successfully describe the distribution of observed
galaxies in the $\Pi_0$ - $i$ diagram
(Figure~\ref{diskpol_model-fig}a). The results in
Table~\ref{model_var-tab} show that for $i \lesssim 50\degr$ the model
is most sensitive to the parameter $f_{\rm B}$. This is so because
wavelength-dependent depolarization such as Faraday dispersion and
differential Faraday rotation are small compared with
wavelength-independent depolarization in the models at 4.8 GHz with a
small inclination. At higher inclination and at lower frequencies,
wavelength-dependent depolarization is more important.  A complete
sample of spiral galaxies observed at 4.8 GHz may be divided into
sub samples by inclination to separate the effects of different physical
parameters. We discuss three inclination ranges.

Galaxies with inclination $i \lesssim 30\degr$ (13.4\% of an unbiased
sample) would display low fractional polarization, provided the galaxy
is axially symmetric. Large-scale departures from axial symmetry
likely result in a higher degree of polarization for these
low-inclination galaxies. A statistical study of the integrated
polarization of face-on spiral galaxies will therefore provide
constraints on the large-scale asymmetry of polarized emission and
therefore magnetic fields in spiral galaxies.  For a sample of face-on
barred galaxies, the orientation of the plane of polarization may be
correlated with the position angle of the bar derived from optical
images.

At inclinations $30\degr \lesssim i \lesssim 50\degr$ (22.3\% of an
unbiased sample), the axially symmetric models are most sensitive to
the ratio of the random magnetic field to the regular magnetic
field.  This inclination range seems most promising to provide
constraints of this parameter for a large sample of
galaxies. Integrated polarimetry at more than one frequency may be
combined to separate wavelength-dependent depolarization from
wavelength-independent depolarization.  More detailed models of the
effect of non-axially symmetric structure such as spiral arms are
required.

At inclinations $i \gtrsim 50\degr$ (64.3\% of an unbiased sample),
the longer path length and larger line-of-sight components of the
regular magnetic field greatly enhance wavelength-dependent Faraday
depolarization and Faraday rotation. This is confirmed by the results
in Table~\ref{model_var-tab}, where the contribution of $B$, $n_e$,
$f_i$, and $2 h$ to the total scatter all rise to approximately equal
amounts at inclination $75\degr$.  The ratio of random to regular
magnetic field is still an important factor, but this can be
constrained by results obtained from galaxies at lower inclination.
Multi-frequency observations would be valuable because the models
suggest that the strongest frequency dependence of depolarization of
the integrated emission occurs for inclinations more than $\sim
60\degr$.

\subsection{Spiral galaxies as polarized background sources}

Are spiral galaxies suitable background sources for rotation measure
studies? Internal Faraday rotation in distant spiral galaxies likely
amounts to rotation measures of a few hundred rad m$^{-2}$.  However,
the correlation between $\thetaB$ and $\thetaopt$ in
Figure~\ref{PA-fig}a, and the models presented in
Section~\ref{model-sec} suggest that the plane of polarization is
related to the orientation of the major axis of the optical disk.  The
polarization angle of an unresolved spiral galaxy is determined by the
orientation of its major axis through the projection of the azimuthal
magnetic field in the galactic disk on the plane of the sky. This is
in part a result of the symmetry of the galaxies, as polarized
emission from areas of the disk opposite to each other with respect to
the center of the galaxy is Faraday rotated by equal but opposite
angles.  Depolarization resulting from differential Faraday rotation
also mitigates the effect of Faraday rotation on the position angle of
the integrated emission.  The correlation in Figure~\ref{PA-fig}a
confirms that the plane of polarization is related to the orientation
of the major axis in real galaxies.

The important implication is that the plane of polarization of a
spiral galaxy that is unresolved by a radio telescope, is independent
of wavelength. The orientation of the plane of polarization can be
determined from the position angle of the optical major axis, although
with a $180\degr$ ambiguity.  As the plane of polarization does not
depend on wavelength, we find the surprising result that {\it
unresolved spiral galaxies behave as idealized polarized background
sources without apparent internal Faraday rotation.} This result
follows from axial symmetry.  The 8.4 GHz polarization data in
Table~\ref{virgo-tab} seem to confirm that the polarization angle of
the integrated emission is independent of wavelength.  More
observations are required to confirm this mainly theoretical result
for more galaxies and at lower frequencies.

The intrinsic position angle of the plane of polarization of an
unresolved spiral galaxy can be estimated from an optical image. It
does not depend on internal Faraday rotation or on the spiral pitch
angle of the magnetic field (Section~\ref{model-sec}).  In principle,
this property allows the detection of possible Faraday rotation by an
intergalactic medium, using a large sample of spiral galaxies.  The
correlation between $\thetaB$ and $\thetaopt$ (Figure~\ref{PA-fig})
can be made for a large number of galaxies, subdivided by distance and
location in the sky. If large-scale magnetic fields exist in the
nearby intergalactic medium, the correlation between $\thetaB$ and
$\thetaopt$ would show an offset for more distant galaxies in a
certain direction. The rotation measure sensitivity of such an
experiment would depend primarily on the sample size of spiral
galaxies with integrated polarimetry, and on the longest wavelength at
which unresolved spiral galaxies remain polarized sources.

\section{Summary and Conclusions}

This paper discusses the integrated polarization properties of nearby
galaxies from archival data at 4.8 GHz, and models for the integrated
polarized emission of a spiral galaxy. We find that

1. Spiral galaxies can show substantially polarized (up to $\sim$20\%
at 4.8 GHz) integrated radio emission. The polarization angle of the
integrated emission of spirals without a (strong) bar is aligned with
the apparent minor axis of the disk.  Barred spirals can also be
polarized sources, but do not show a clear correlation of polarization
angle with the orientation of the major axis.

2. The highest degree of polarization, and the largest spread in
fractional polarization, are found for spiral galaxies with
intermediate to high inclination ($i \gtrsim 50\degr$). 

3. A sub-sample of nine spiral galaxies in the Virgo cluster does not
show a significant difference in integrated polarization properties
from the field galaxies in our sample. This suggests that physical
distortions of these galaxies are not so strong that they have a
noticeable effect on the integrated polarization. Confirmation by a
larger, complete sample is required.

4. The fractional polarization of integrated emission is smaller for
galaxies with a higher radio luminosity. This result suggests a
difference in properties of the magnetic field and/or Faraday
depolarization in luminous spiral galaxies.

5. Our models for the integrated polarization of axially symmetric
spiral galaxies cover the same area in the $\Pi_0$-$i$ diagram
(Figure~\ref{diskpol_model-fig}) as the observed galaxies
without a strong bar. The models indicate that the main cause of
variation in the fractional polarization of galaxies with inclination
less than $50\degr$ is the ratio of the random to the regular magnetic
field component in the disk.

6. Our models provide a probability density function of fractional
polarization of spiral galaxies as a function of wavelength that can
be used to predict the number of spiral galaxies in deep polarization
surveys, and to study the statistical properties of spiral galaxies in
such surveys.

7. The integrated polarization of spiral galaxies opens the
possibility to study magnetic field properties and Faraday rotation in
large samples of spiral galaxies, and in galaxies at high redshift.  A
deep 1 - 2 GHz survey with the SKA could detect normal spiral galaxies
at redshift 1 or higher \citep{wilman2008} with a total intensity flux
$\sim 100\ \mu \rm Jy$. These can be compared in a
resolution-independent way with integrated polarization measurements
of low-redshift galaxies at the same restframe frequency.

8. Unresolved symmetric spiral galaxies behave as idealized background
sources without internal Faraday rotation. They can be used to detect
large-scale magnetic fields in the intergalactic medium, e.g. with the
SKA \citep{gaensler2004, stepanov2008}.

\acknowledgements We thank Marek We\.zgowiec and Marek Urbanik for
giving us the maps of the Virgo galaxies, some of them prior to
publication. The authors thank K. T. Chy\.zy for the use of data for
NGC~4254 and NGC~4736. We also acknowledge Maike Lensing for her help
to prepare the radio maps for the flux integration. The authors thank
the referee for comments that helped to improve the manuscript.  This
research has made use of the NASA/IPAC Extragalactic Database (NED)
which is operated by the Jet Propulsion Laboratory, California
Institute of Technology, under contract with the National Aeronautics
and Space Administration.

\end{document}